\documentclass[
showpacs,
floatfix,
aps,
prb,
twocolumn,
superscriptaddress,
amssymb,
]{revtex4-1}

\usepackage{times}
\usepackage{amssymb}
\usepackage{latexsym}
\usepackage[dvips]{graphicx}
\usepackage{amsmath}
\usepackage{graphicx}
\usepackage{dcolumn}
\usepackage{amsfonts}
\usepackage{bm}
\usepackage{epsfig}

\newcommand{\be}{\begin{equation}}
\newcommand{\ee}{\end{equation}}
\newcommand{\bea}{\begin{eqnarray}}
\newcommand{\eea}{\end{eqnarray}}

\def\oc{\omega_{\mbox{\scriptsize {c}}}}

\def\lp{\left (}
\def\rp{\right )}

\newcommand{\req}[1]{Eq.\,(\ref{#1})}
\newcommand{\rEq}[1]{Equation\,(\ref{#1})}

\newcommand{\rfig}[1]{Fig.\,\ref{#1}}

\newcommand{\rfigs}[2]{Figs.\,\ref{#1},\,\ref{#2}}
\newcommand{\rref}[1]{Ref.\,\onlinecite{#1}}
\newcommand{\rrefs}[2]{Refs.\,\onlinecite{#1},\,\onlinecite{#2}}

\def\bpar{B_\parallel}
\def\bper{B_\perp}

\def\gpar{g_\parallel}
\def\gper{g_\perp}

\def\apar{a_\parallel}
\def\aper{a_\perp}

\def\ds{\Delta_{\mbox{\scriptsize {s}}}}

\def\dz{\Delta_Z}
\def\dx{\Delta_X}

\def\az{\alpha_Z}
\def\ap{\beta}
\def\am{\alpha_\gamma}

\def\ax{\alpha_X}
\def\dodd{\Delta_{\rm odd}}
\def\deven{\Delta_{\rm even}}

\def\lp{\left (}
\def\rp{\right )}

\def\nus{\nu_{\mbox{\scriptsize {s}}}}
\def\nuc{\nu_{\mbox{\scriptsize {c}}}}

\def\aa{$\theta = 87.00^\circ\,(\gamma = 19.1)$}
\def\ab{$\theta = 87.43^\circ\,(\gamma = 22.3)$}
\def\ac{$\theta = 87.76^\circ\,(\gamma = 25.6)$}
\def\ad{$\theta = 87.98^\circ\,(\gamma = 28.4)$}

\def\ba{$\theta = 88.04^\circ\,(\gamma = 29.2)$}
\def\bb{$\theta = 88.16^\circ\,(\gamma = 31.1)$}
\def\bc{$\theta = 88.46^\circ\,(\gamma = 37.2)$}

\def\ca{$\theta = 88.84^\circ\,(\gamma = 49.6)$}
\def\cb{$\theta = 89.02^\circ\,(\gamma = 58.1)$}
\def\cc{$\theta = 89.09^\circ\,(\gamma = 62.7)$}

\def\mob{$\mu \approx 2.4 \cdot 10^7$ cm$^2$/Vs~}
\def\den{$n_e \approx 2.85 \cdot 10^{11}$ cm$^{-2}$~}

\begin{document}
\title{ 
Shubnikov-de Haas oscillations in GaAs quantum wells in tilted magnetic fields
}
\author{A.\,T. Hatke}
\affiliation{School of Physics and Astronomy, University of Minnesota, Minneapolis, Minnesota 55455, USA}

\author{M.\,A. Zudov}
\email[Corresponding author: ]{zudov@physics.umn.edu}
\affiliation{School of Physics and Astronomy, University of Minnesota, Minneapolis, Minnesota 55455, USA}

\author{L.\,N. Pfeiffer}
\affiliation{Department of Electrical Engineering, Princeton University, Princeton, New Jersey 08544, USA}

\author{K.\,W. West}
\affiliation{Department of Electrical Engineering, Princeton University, Princeton, New Jersey 08544, USA}


\begin{abstract}
We report on quantum magneto-oscillations in an ultra-high mobility GaAs/AlGaAs quantum well at very high ($\ge 87^\circ$) tilt angles.
Unlike previous studies, we find that the spin and cyclotron splittings become equal over a {\em continuous range} of angles, but only near certain, angle-dependent filling factors.
At high enough tilt angles, Shubnikov-de Haas oscillations reveal a prominent beating pattern, indicative of {\em consecutive} level crossings, {\em all} occurring at the same angle.
We explain these unusual observations by an in-plane field-induced increase of the carrier mass, which leads to accelerated, filling factor-driven crossings of spin sublevels in tilted magnetic fields.
\end{abstract}
\maketitle

When a two-dimensional electron system (2DES) is subject to a perpendicular magnetic field $\bper$, its low temperature magnetoresistivity exhibits well-known Shubnikov-de Haas oscillations (SdHO).
SdHO originate from the magnetic quantization of the energy spectrum, 
\be
E_{N,\mp} = \hbar\oc \lp N + 1/2 \rp \pm \ds/2\,,
\label{eq.sp}
\ee
where $\hbar\oc=\hbar e \bper/m^\star$ is the cyclotron energy, $m^\star$ is the effective mass, $N$ is the Landau level index, and $\ds$ is the sum of the single particle Zeeman energy $\dz$ and the exchange energy $\dx$.
The resistivity of the 2DES senses the density of electron states at the Fermi level, $\varepsilon_F$, and will show a minimum whenever $\varepsilon_F$ falls in the middle between the energy levels, see \req{eq.sp}, i.e., when the filling factor $\nu=2\varepsilon_F/\hbar\oc$ acquires an integer value.

The strength of the SdHO at odd and even $\nu$ is determined by {\em spin splitting}, $\dodd = \ds$, and {\em cyclotron splitting}, $\deven = \hbar\oc - \ds$, respectively.
Furthermore, the exchange contribution to the spin splitting becomes important only when there is an appreciable population difference between $(N,+)$ and $(N,-)$ sublevels.
The latter turns off rather abruptly at some critical, disorder-dependent filling factor $\nus$.
The associated disappearance of the odd-$\nu$ SdHO minima is referred to as a critical collapse of the exchange-enhanced spin splitting.\citep{fogler:1995,leadley:1998,piot:2005,piot:2007,pan:2011} 
At odd $\nu \gg 1$, but considerably lower than $\nus$, the exchange energy scales with the cyclotron energy  $\dx=\ax\hbar\oc$,\citep{fogler:1995,aleiner:1995} while at even $\nu$, one can set $\dx \approx 0$ and $\ds \approx \dz = \az\hbar\oc$.
Since in GaAs $\az \ll 1$, $\ax < 1$,\citep{note:5} and, most importantly, since there are no critically collapsing contributions to $\deven$, the SdHO at even-$\nu$ persist to filling factors much larger than $\nus$.\citep{note:6}

Adding an in-plane magnetic field will increase the Zeeman energy since the latter scales with the total field.
On the other hand, one usually assumes that both the cyclotron and the exchange energies depend only on $\bper$.\citep{leadley:1998,piot:2007,note:8}
As a result, when the magnetic field is tilted by angle $\theta$ away from the sample normal, for any given $\nu$, the cyclotron splitting, $\deven \propto 1 - \az/\cos\theta$, will decrease with $\theta$ and eventually approach the increasing spin splitting, $\dodd \propto \ax + \az/\cos\theta$. 
This approach forms a basis of the {\em coincidence method},\citep{fang:1968} which was successfully used to study the exchange contribution.\citep{nicholas:1988,leadley:1998}
The ``coincidences'', manifested as equally strong even-$\nu$ and odd-$\nu$ SdHO, will occur when\citep{note:7} $\deven = \dodd$ or
\be
\gamma_i \equiv \frac 1 {\cos \theta_i} = \frac {i - \ax}{2\az}\,,~~i=1,3,5,...\,.
\label{eq.old}
\ee
\rEq{eq.old},\citep{note:3} dictating that the coincidences should occur {\em simultaneously} for all $\nu$, was confirmed experimentally\citep{nicholas:1988,leadley:1998} for $i = 1$, $\theta_1 \approx 87.2^\circ$ $(\gamma_1 \approx 20)$.  

In this Letter we report on spin-resolved SdHO in an ultra-high mobility GaAs/AlGaAs quantum well at very high tilt angles ($\theta \ge 87^\circ$).
We find that the evolution of the SdHO waveform depends sensitively not only on $\theta$, as predicted by \req{eq.old}, but also on the filling factor. 
More specifically, we find that coincidence conditions, $\deven = \dodd$, can be satisfied {\em over a range of tilt angles}, but only at certain filling factors.
These filling factors monotonically increase with $\theta$ and eventually diverge at $\theta=\theta_i$. 
At $\theta > \theta_1$, SdHO reveal a beating pattern indicating {\em consecutive} ($i = 3, 5$), filling factor-driven coincidences, occurring at the same $\theta$.
These findings are in contrast with \req{eq.old} and previous experiments.\citep{nicholas:1988,leadley:1998}
To explain this unusual behavior, we propose a model which takes into account an in-plane field-induced increase of the effective mass due to the finite width of our 2DES.
This increase leads to a non-monotonic dependence of the cyclotron splitting on $\nu$ and its eventual collapse resulting in accelerated crossings of spin sublevels.

Our sample is a Hall bar (width 200 $\mu$m) fabricated from a 29 nm-wide GaAs-Al$_{0.24}$Ga$_{0.76}$As quantum well, grown by molecular beam epitaxy, having density \den and mobility \mob. 
The resistivity was recorded using a standard low-frequency (a few Hertz) lock-in technique at temperature $T \approx 0.3$ K in sweeping magnetic field.
The tilt angle $\theta$, controlled {\em in situ} by a precision rotator, was held constant during each magnetic field sweep. 

\begin{figure}[t]
\includegraphics{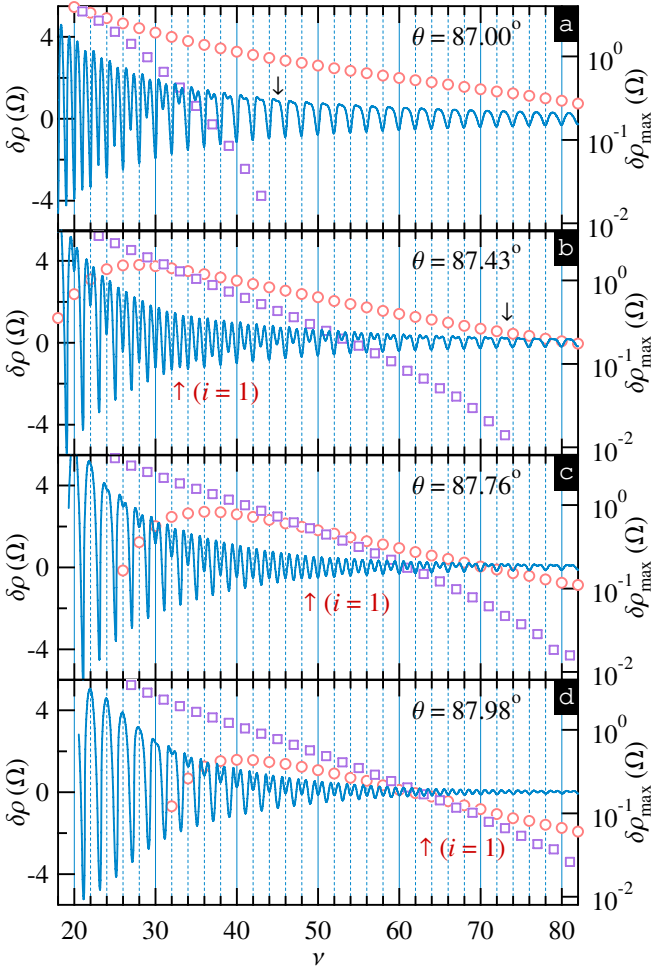}
\vspace{-0.1 in}
\caption{(Color online)
Oscillatory part of the resistivity $\delta\rho$ (left axis) and amplitude $\delta\rho_{\max}$ (right axis) versus the filling factor $\nu$ measured at tilt angles (a) \aa, (b) \ab, (c) \ac, and (d) \ad. 
Even-$\nu$ (circles) and odd-$\nu$ (squares) amplitudes become equal at $\nu = \nu_1$, as marked by $\uparrow$.
}
\vspace{-0.15 in}
\label{fig1}
\end{figure}

In \rfig{fig1} we present the oscillatory part of the resistivity, $\delta\rho$, (left axes), obtained from the total resistivity by subtracting a slowly varying background, versus $\nu$ at different $\theta$, as marked. 
The corresponding amplitudes, $\delta\rho_{\max}$, at even and odd filling factors are shown on the right (logarithmic) axes as circles and squares, respectively.
These amplitudes were evaluated from $\delta\rho$ at the minima and the average of the neighboring maxima. 

At the lowest $\theta$, \rfig{fig1}(a), we observe that at $\nu\gtrsim 45$ the SdHO minima occur only at even $\nu$.
The corresponding (even-$\nu$) amplitude follows roughly exponential dependence, as expected from the Lifshitz-Kosevich formula.\citep{isihara:1986}
The minima at odd filling factors first appear at $\nu = \nus \approx 45$ (cf. $\downarrow$), which marks the onset of the (exchange enhanced) spin splitting.
At $\nu<\nus$, the odd-$\nu$ amplitude increases rapidly and approaches the even-$\nu$ amplitude.
Overall, the data in \rfig{fig1}(a) are very similar to the data obtained at all lower $\theta$, including $\theta = 0$, the only difference being that $\nus$ monotonically increases with $\theta$.
Such an increase of $\nus$ is a result of accelerated opening of the spin gap due to the growing single-particle Zeeman energy,\citep{note:4} and is well established experimentally.\citep{leadley:1998,piot:2007} 

At the next tilt angle, \rfig{fig1}(b), in addition to the onset of spin splitting moving to much higher filling factors, $\nus \approx 73$, the data reveal other interesting features, some of which are rather unexpected. 
First, the amplitudes at odd-$\nu$ and even-$\nu$ become equal at $\nu_1 \approx 33$ (cf. $\uparrow$).
Second, at lower filling factors, $\nu < \nu_1$, the odd-$\nu$ amplitude continues to grow but the even-$\nu$ amplitude first saturates and then starts to {\em decrease}.
As evidenced by the data at higher $\theta$, \rfig{fig1}(c), this decrease leads to a complete disappearance of the even-$\nu$ amplitude at $\nu = \nuc \approx 24$.
The data at still higher tilt, \rfig{fig1}(d), confirm this observation and let us conclude that {\em all} characteristic filling factors, $\nus$, $\nu_1$, and $\nuc$, increase with $\theta$. 

The observed amplitude crossings indicate that at $\nu = \nu_1$ the spin splitting becomes equal to the cyclotron splitting.
On the other hand, the vanishing of the even-$\nu$ amplitude at $\nu=\nuc$ signals a collapse of the cyclotron gap, which is a precursor of crossing of spin sub-levels from the neighboring Landau levels.
Such crossing would occur when $\deven = \hbar\oc - \dz = 0$, so the observed collapse indicates that, with increasing $1/\nu$, $\dz$ is growing faster than $\hbar\oc$. 
As we show below, the collapse of $\deven$ can be explained by a $\bpar$-induced increase of the electron effective mass, which leads to a sublinear dependence of $\hbar\oc$ on $1/\nu$ and its eventual saturation.

We further notice that, since $\nus - \nu_1 \gg 1$, the exchange energy at odd filling factors should be close to its maximum value, $\ax \hbar \oc$.\citep{aleiner:1995,fogler:1995,leadley:1998,piot:2007} 
This conclusion is also supported by the observations that, at $\nu \lesssim \nu_1$, the separation between spin-split maxima is close to unity, $\delta\nu \approx 1$, and that the odd-$\nu$ amplitude shows excellent exponential dependence.
Finally, at all $\theta \gtrsim 87^\circ$, $\gamma\az$ is no longer a small fraction and $\dz$ gives a significant (if not dominant) contribution to the spin splitting. 
\begin{figure}[t]
\includegraphics{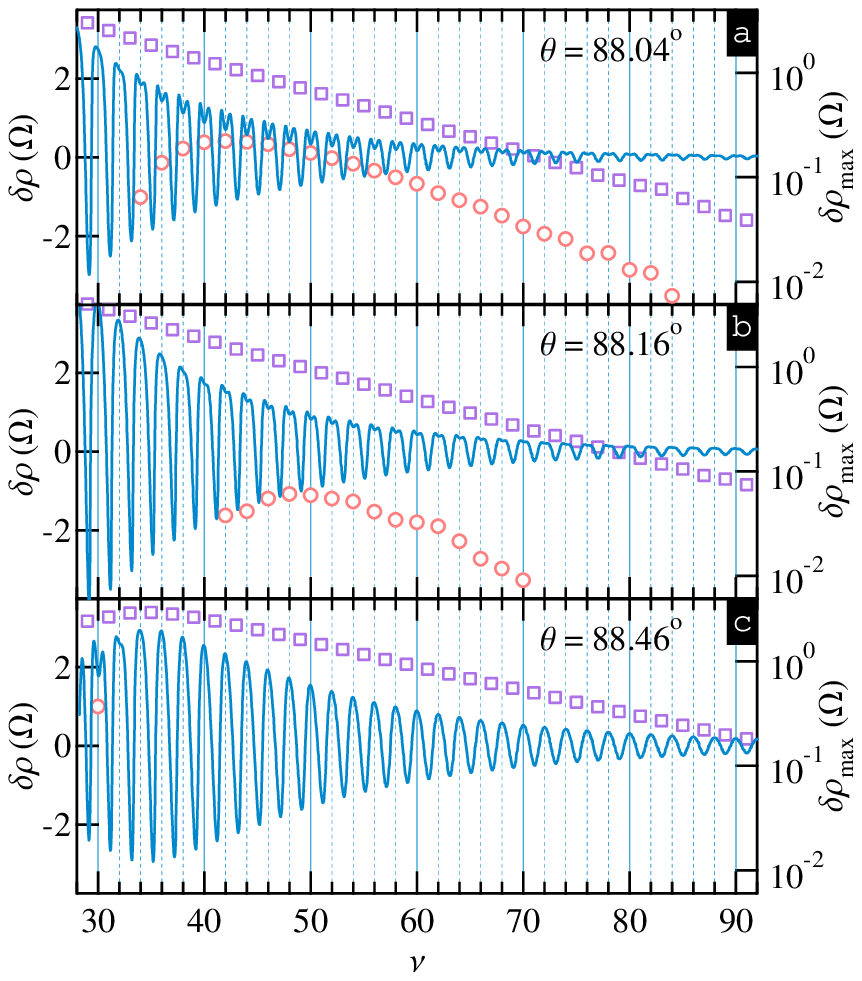}
\vspace{-0.1 in}
\caption{(Color online)
Same as \rfig{fig1} (a) \ba, (b) \bb, (c) \bc.
}
\vspace{-0.15 in}
\label{fig2}
\end{figure}

SdHO measured at higher tilt angles (as marked) are shown in \rfig{fig2}, which has the same layout as \rfig{fig1}.
Compared to the data at lower $\theta$ discussed above, \rfig{fig2} reveals further significant changes of the SdHO waveform.
Here, we observe that the even-$\nu$ amplitude remains considerably smaller than the odd-$\nu$ amplitude over the whole range of $\nu$ and the crossing never takes place.
Furthermore, we observe in \rfig{fig2}(c) that the even-$\nu$ SdHO disappear completely, indicating that the condition for the first level crossing is approximately satisfied for all filling factors under study.
However, closer examination of \rfig{fig2}(c) reveals a maximum of the odd-$\nu$ amplitude at $\nu \approx 35$ and the re-emergence of the even-$\nu$ SdHO to the left of this maximum, indicating that at these filling factors the crossing of spin sublevels has already occurred.

\begin{figure}[t]
\includegraphics{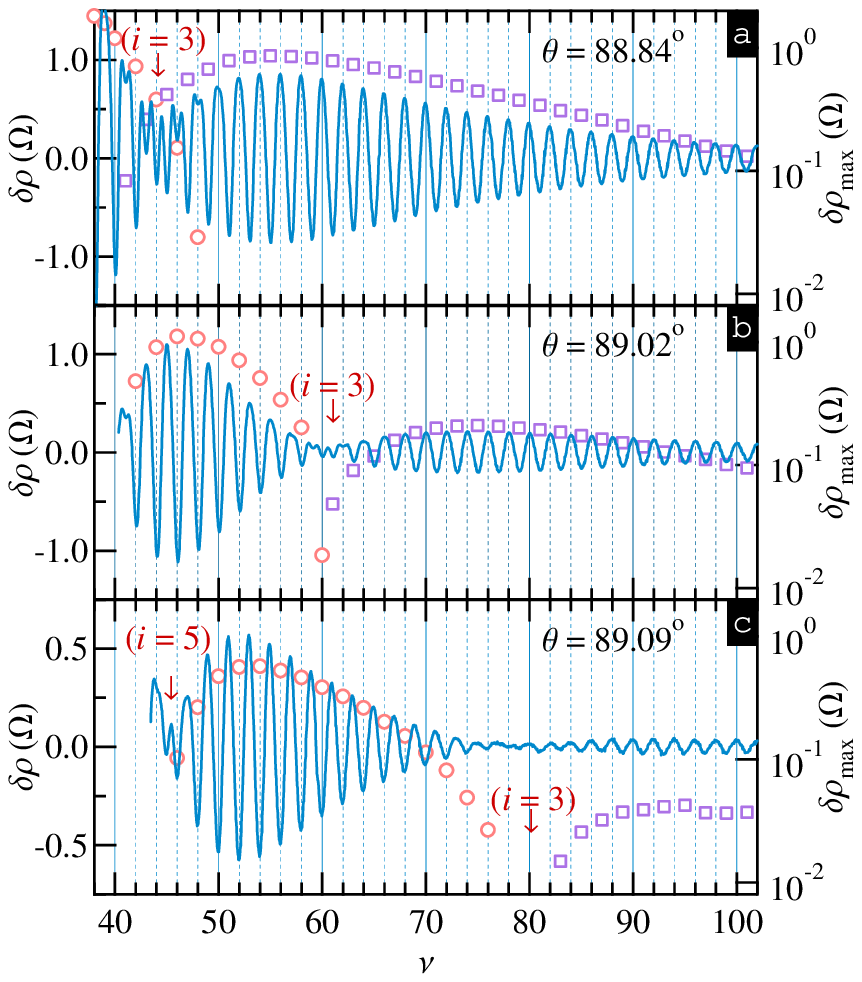}
\vspace{-0.1 in}
\caption{(Color online)
Same as \rfigs{fig1}{fig2} (a) \ca, (b) \cb, (c) \cc.
}
\vspace{-0.15 in}
\label{fig3}
\end{figure}

In \rfig{fig3} we present the data obtained at still higher tilt angles.
The most striking feature of all three data sets is a node which occurs at $\nu = \nu_3$ (cf. $\downarrow$). 
Similar to the ``coincidence'' at $\nu=\nu_1$, this node moves to progressively higher $\nu$ with increasing $\theta$.
Passing through the node from high to low filling factors, odd-$\nu$ SdHO decay away while the even-$\nu$ SdHO set in.
As a result, at $\nu \approx \nu_3$ the amplitudes of the odd-$\nu$ and even-$\nu$ SdHO again become close to each other, reflecting equal gaps, similar to the situation at $\nu \approx \nu_1$ (cf. \rfig{fig1}).

While at $\nu < \nu_3$, even-$\nu$ SdHO dominate the response, the oscillation amplitude is not monotonic, and after passing through a maximum again decays away [cf. \rfig{fig3}(b)].
As shown by the data at the highest tilt angle, \rfig{fig3}(c), this decay gives rise to another node, at $\nu = \nu_5 \approx 45$, which signals yet another crossing of the even-$\nu$ and odd-$\nu$ amplitudes.
While the oscillation patterns shown in \rfig{fig3} might appear very different from those shown in \rfig{fig1} and \rfig{fig2}, as we show next, all observations can be attributed to subsequent level crossings.
Clearly, such crossings cannot be described by \req{eq.old}, which is independent of $\nu$.

To explain our findings we propose a model based on finite thickness effects which, in combination with $\bpar$, lead to an increase of the effective mass.
This increase was studied theoretically\citep{khaetskii:1983,maan:1984,merlin:1987,tang:1988,smrcka:1990,smrcka:1994} and confirmed in experiments examining the temperature damping of the SdHO amplitude\citep{smrcka:1995} and the shift of magnetoplasmon resonances.\citep{batke:1986,kozlov:2011}
Following \rref{kozlov:2011}, we use the expression for a parabolic confining potential\citep{khaetskii:1983,smrcka:1994} and substitute the intersubband splitting of our quantum well, $\varepsilon_{10}$, for the intersubband spacing in the parabolic well. 
At $\gamma \gg 1$, the effective mass increases with respect to its value at $\theta = 0$ by a factor $1/\am$, which is given by\citep{khaetskii:1983,smrcka:1994}
\be
\frac 1 {\am}= \frac {m^\star_\theta}{m^\star_0}  
\approx \sqrt{1+\lp \frac {\ap\gamma}{\nu}\rp^2}\,,
\label{eq.mass}
\ee
where $\ap = 2 \varepsilon_F/\varepsilon_{10}$ and $\varepsilon_F$ is evaluated at $\theta = 0$.
Since $\ap$ is of the order of unity,\citep{note:1} a significant mass increase is expected at $\nu \lesssim \gamma$.
This increase will result in a sublinear dependence of the cyclotron energy on $1/\nu$ and its eventual saturation at $\hbar\oc = \varepsilon_{10}/\gamma \ll \varepsilon_{10}$, in the limit of $\nu \ll \gamma$.
As a result, at $\theta \neq 0$, the cyclotron splitting, $\deven \equiv \hbar\oc - \dz \propto [\am(\nu) - \az\gamma]/\nu$, is no longer a monotonic function of $\nu$ and, after reaching a maximum,
 will decrease and eventually collapse.
This non-monotonic behavior of the cyclotron splitting translates to the $\nu$-dependence of the even-$\nu$ SdHO amplitude which is observed in \rfig{fig1}(b)-(d) and \rfig{fig2}(a)-(b).

Coincidence conditions, which take into account the mass increase, \req{eq.mass}, can now be obtained by replacing $i$ with $i\am$ in \req{eq.old} and solving for $\nu$, which results in
\be
\nu_i \approx \frac {\beta\gamma}{\sqrt{i^2/\lp 2\az\gamma+\ax\rp^{2} - 1}}.
\label{eq.new}
\ee
In the limit of large $\nu$, $\am = 1$ and \req{eq.new} reduces to \req{eq.old}, which defines $\nu$-independent coincidence angles.

Tilting the sample with respect to the magnetic field is also known to affect $\az$ due to the anisotropy of the $g$-factor.
Recent experiments on electron-spin resonance in a symmetric quantum well, similar to ours, have shown that $g_0 = \sqrt{\gper^2 \cos^2\theta + \gpar^2 \sin^2\theta}$, where $\gper = 0.414$ and $\gper = 0.340$.\citep{nefyodov:2011b}
At $\gamma \gg 1$, we obtain $g_0 \approx \gpar = 0.340$.
It is also known that at high magnetic fields, quadratic-in-$B$ corrections to the Zeeman energy become important.\citep{dobers:1988,nefyodov:2011a,nefyodov:2011b}
These terms can be incorporated into the $g$-factor which acquires linear $B$ dependence, $g = g_0 - a B$.
At $\theta \neq 0$, $a \approx (\cos\theta/g)(\gper\aper\cos^2\theta+\gpar\apar\sin^2\theta)$, where $\aper \sim \apar \sim 0.01$ T$^{-1}$.\citep{nefyodov:2011b}
At $\gamma \gg 1$, $a \approx \apar/\gamma \ll \apar$, and the correction to the $g$-factor, $aB$, remains less than 1\% in our experiment even at the highest magnetic fields.
We will therefore assume a $B$- and $\theta$-independent $g$-factor, $g = g_0 = 0.34$, in our analysis.

\begin{figure}[t]
\includegraphics{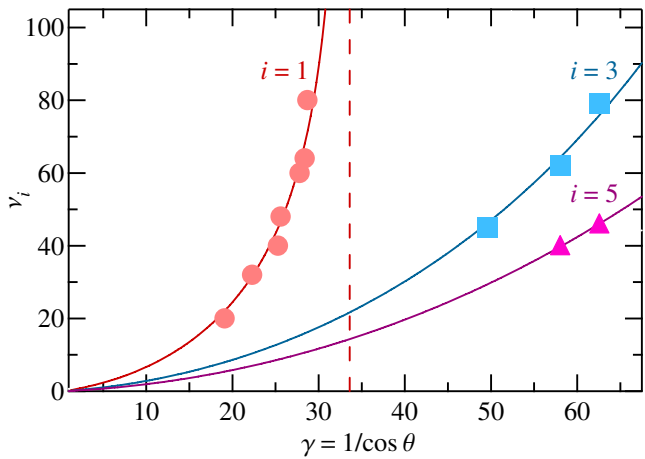}
\vspace{-0.1 in}
\caption{(Color online)
Filling factors $\nu_1$ (circles), $\nu_3$ (squares), and $\nu_5$ (triangles), corresponding to crossings of the even-$\nu$ and odd-$\nu$ amplitudes as functions of $\gamma$.
Solid lines are fits to the data (see text).
Dashed vertical line is the asymptote of the $i=1$ coincidence at large $\nu$.
}
\vspace{-0.15 in}
\label{fig4}
\end{figure}
To demonstrate that our experimental findings can be described by \req{eq.new}, we present in \rfig{fig4} the filling factors $\nu_1$ (circles), $\nu_3$ (squares), and $\nu_5$ (triangles), corresponding to crossings of the even-$\nu$ and odd-$\nu$ amplitudes as functions of $\gamma$.
Solid lines represent fits to the data using \req{eq.new} with $\az \equiv g_0 m^\star/2m = 0.0114$, $\ax = 0.234$,\citep{note:5} and $\beta$ used as a fitting parameter.
For $\nu_1$, the fit generates $\beta \approx 1.3$, which is close to our estimated value of $\beta = 1.1$.\citep{note:1}
The vertical dashed line is drawn at $\theta=\theta_1$, calculated using \req{eq.old}.
For $\nu_3$ and $\nu_5$, the fits produce $\beta \approx 1.8$ and $\beta \approx 2.1$, respectively.
Although not unreasonable, these values are higher than our estimate, which likely reflects the limitations of our model.

In summary, we have demonstrated that in a typical ultra-high mobility GaAs/AlGaAs quantum well the evolution of Shubnikov-de Haas oscillations in tilted magnetic fields cannot be described by a simple series of level crossings characterized by a set of angles defined by \req{eq.old}. 
Instead, multiple level crossings can be realized for a given angle, but only near some characteristic filling factors, see \req{eq.new} and \rfig{fig4}.
These characteristic filling factors depend sensitively on $\theta$, diverging near $\theta = \theta_i$ given by \req{eq.old}.
Both multiple level crossings and filling factor-driven collapses of the energy gaps can be explained by finite thickness effects, which give rise to the effective mass increase induced by in-plane magnetic field.

We thank I. Dmitriev, M. Khodas, T. Shahbazyan, B. Shklovskii, S. Studenikin, and M. Dyakonov for discussions and G. Jones and T. Murphy for technical assistance.
This work was supported by the US Department of Energy, Office of Basic Energy Sciences, under Grant No. DE-SC002567. 
The work at Princeton was partially funded by the Gordon and Betty Moore Foundation and by the NSF MRSEC Program through the Princeton Center for Complex Materials (DMR-0819860).
A portion of this work was performed at the National High Magnetic Field Laboratory, which is supported by NSF Cooperative Agreement No. DMR-0654118, by the State of Florida, and by the DOE.


\begin{thebibliography}{28}
\bibitem[{not({\natexlab{a}})}]{note:5}
\bibinfo{note}{According to \rref{aleiner:1995}, $\ax = \ln(k_Fa_B)/\pi k_F
  a_B$, where $k_F$ is the Fermi wave number, and $a_B$ is the effective Bohr
  radius. In our 2DES we estimate $\ax \approx 0.23$.}
\bibitem[{not({\natexlab{d}})}]{note:7}
\bibinfo{note}{\rEq{eq.old} for $i = 1$ coincides with Eq. (11) of
  \rref{leadley:1998} if $\ax$ is replaced by 2$\ax$. The extra factor appears
  because in \rref{leadley:1998} $\dx$ is included in both $\dodd$ and
  $\deven$.}
\bibitem[{not({\natexlab{e}})}]{note:3}
\bibinfo{note}{\rEq{eq.old} should also be approximately valid at even $i$,
  corresponding to level crossings, if one takes the exchange part to be only
  partially ($\simeq 50 \%$) enhanced (\rref{nicholas:1988}).}
\bibitem[{not({\natexlab{g}})}]{note:1}
\bibinfo{note}{Using infinite square well approximation for our 2DES gives
  $E_{10} \approx 225$ K and $\ap \approx 1.1$ ($E_F \approx 125$ K).}
\end{thebibliography}

\end{document}